\documentclass[aps,prl,twocolumn,superscriptaddress,floatfix,showpacs]{revtex4}

\usepackage{float}
\usepackage{placeins}
\usepackage{amssymb,amsmath}
\usepackage{graphicx}
\usepackage[colorlinks=true,linkcolor=blue,citecolor=blue,urlcolor=blue]{hyperref}

\begin{document}

\title{Differential-thermal analysis around and below the critical temperature $T_c$ of various low-$T_c$ superconductors: A comparative study}

\author{M.~Reibelt}
\email[]{reibelt@physik.uzh.ch}
\author{A.~Schilling}
\affiliation{Physik-Institut University of Zurich, Winterthurerstr.
$190$, CH-$8057$ Zurich, Switzerland}
\author{P.~C.~Canfield}
\affiliation{Ames Laboratory, Department of Physics and Astronomy,
Iowa State University, Ames, Iowa $50011$, USA}
\author{G.~Ravikumar}
\affiliation{Technical Physics and Prototype Engineering Division, Bhabha
Atomic Research Centre, Mumbai-$400085$, India}
\author{H.~Berger}
\affiliation{Institut de Physique de la Mati$\grave{e}$re Complexe, Ecole Polytechnique F$\acute{e}$d$\acute{e}$rale de Lausanne, CH-$1015$ Lausanne, Switzerland}
\date{\today}

\begin{abstract}
We present specific-heat data on the type-II superconductors V$_3$Si, LuNi$_2$B$_2$C and NbSe$_2$ which were acquired with a low-temperature thermal analysis (DTA) technique. We compare our data with available literature data on these superconductors. In the first part we show that the DTA technique allows for fast measurements while providing a very high resolution on the temperature scale. Sharp features in the specific heat such as at the one at the transition to superconductivity are resolved virtually without instrumental broadening. In the second part we investigate the magnetic-field dependence of the specific heats of V$_3$Si and LuNi$_2$B$_2$C at a fixed temperature $T=7.5\, $K to demonstrate that DTA techniques also allow for sufficiently precise measurements of absolute values of $c_p$ even in the absence of a sharp phase transition. The corresponding data for V$_3$Si and LuNi$_2$B$_2$C are briefly discussed.
\end{abstract}


\maketitle

\section{Introduction}

The specific heat is a bulk thermodynamic quantity determined uniquely for every material by its spectrum of excitations. The measurement of the specific heat is a basic technique to reveal the physical properties of a material because it can, in principle, be calculated \emph{ab initio} from a suitable physical model. The prediction of a linear specific heat, for example, is one of the most important consequences of Fermi-Dirac statistics for electrons in a metal, and its measurement provides a simple test to the electron gas theory of a metal. Because the specific heat is a quantity that is probing the whole sample volume, it is significant to characterize volume effects such as the superconducting state in a material. Therefore, specific-heat measurements have traditionally been of great importance for investigations in the field of superconductivity. The electronic specific heat of superconductors can be expected to yield information about the nature of the superconducting state; and its temperature ($T$) dependence should in particular be related to the energy gap \cite{Cooper1959May}. Specific-heat measurements can also provide information about fluctuation effects \cite{Thouless1975April,Rao1981February,Zally1971December,Park2002,Lortz2006September}, and they are particularly suited to measure phase transitions and to explore the phase diagram of superconductors. Improvements in the techniques for specific-heat measurements are therefore of special value. As an example, the first-order nature of the melting of the flux-line lattice in high-temperature superconductors in the mixed state has been unambiguously proven for the first time by high-resolution specific-heat measurements based on a differential-thermal analysis (DTA) method developed for this purpose \cite{Schilling1995October,Schilling1996August}. The DTA method used in our laboratory is particularly suited to detect sharp phase transitions, e.g., first-order phase transitions, and it does not require large sample masses. Since high-quality samples often only exist in form of small crystals, such a sensitive method is of particular interest. In addition, it is often desirable to choose a method that yields a high data-point density within a reasonable measuring time. Common techniques, such as standard heat-pulse or relaxation methods, are sensitive but very time consuming. As an example, the relaxation method implemented in a commercial PPMS platform (Quantum Design) usually takes several \emph{minutes} per data point for data acquisition, while the DTA technique takes only approximately $3\, $s per data point, at an impressive data-point density of typically $170$ data points per Kelvin. High resolution and high data-point density are needed for the observation of small and sharp effects. In the first part of this work we are comparing corresponding DTA measurements on V$_3$Si, LuNi$_2$B$_2$C, and NbSe$_2$ with measurements done in a commercial PPMS platform and with data from the literature, with a special focus on the transition width of the normal to the superconducting state.\\
\indent Historically, specific-heat ($c_p$) measurements using DTA techniques have been proven to be particularly powerful when studying relative changes in $c_p$ at sharp phase transitions. However, to extract exact absolute values, or to identify smooth, featureless trends (such as the exact functional temperature dependence of $c_p$), is not easy using this technique due to various reasons \cite{Schilling1995October}. In superconductors, not only the discontinuity in the specific heat at $T_c$ contains valuable information about the superconducting state, but also the temperature and the field dependence of this quantity below $T_c$ are of utmost interest. From the temperature dependence $c_p(T)$ one can draw conclusions about zeroes in the gap function, which may help clarify questions about the nature of the order parameter \cite{Goll2005December}. A power-law dependence of $c_p$ on temperature has been observed, for example, in YBa$_2$Cu$_3$O$_7$ \cite{Wright1999February}, with a $T^2$ term for a magnetic field $H=0$ and an $H^{1/2}T$ term for $H \neq 0$ and low $T$, with a crossover to a stronger $T$ dependence at high $T$. Peculiar $T$ dependencies, very distinct from the standard one-band \emph{s}-wave exponential law, can also be expected in multigap superconductors, as has been measured for example in MgB$_2$ \cite{Bouquet2002December}. On the other hand, the field dependence of $c_p(H)$ at constant temperature may also be influenced by the symmetry of the order parameter and also by other factors \cite{Moler1994November,Wright1999February,Volovik1993,Ramirez1995February}. In the second part of this manuscript we show that specific-heat data from DTA measurements can, to some extent, also give useful information on absolute values of $c_p$, and in particular reveal trends even in the absence of a sharp phase transition. We briefly discuss our data on the magnetic field dependence of $c_p(H)$  for V$_3$Si, and LuNi$_2$B$_2$C at a fixed temperature $T=7.5\, $K.\\

\section{Sample Characterization}

\begin{figure}
\includegraphics[width=75mm,totalheight=200mm,keepaspectratio]{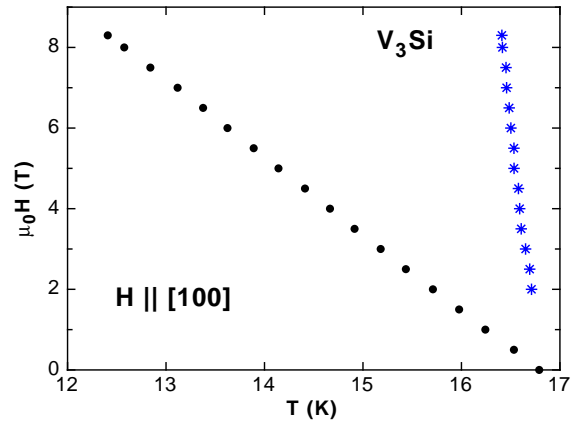}
\caption{Phase diagram of the investigated V$_3$Si single crystal, showing the upper critical field $H_{c2}(T)$ (filled circles), and the martensitic transition $H_M(T)$ (stars). The magnetic field was oriented parallel to the $[100]$ direction.
\label{fig.physicaC21}}
\end{figure}

\begin{figure}
\includegraphics[width=75mm,totalheight=200mm,keepaspectratio]{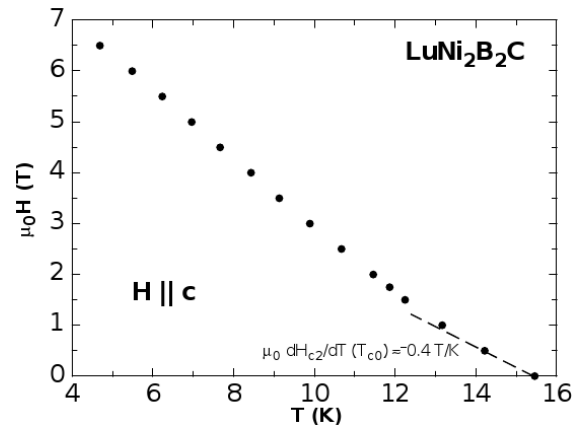}
\caption{Upper critical field $H_{c2}(T)$ of the investigated LuNi$_2$B$_2$C single crystal. The magnetic field was oriented perpendicular to the layers. Note the unusual positive curvature of $H_{c2}(T)$ near $T_{c0}$.
\label{fig.physicaC20}}
\end{figure}

The V$_3$Si single crystalline sphere (mass $m \approx 11.1\, $mg) used for this study has been previously characterized using magnetization measurements by K\"upfer \emph{et~al.}~(sample "SA" in \cite{Kupfer2004}). The authors irradiated V$_3$Si samples with fast neutrons, and the present sample was further annealed for $2\, $h at $630\, ^\circ$C in order to decrease pinning effects. We determined the transition temperature to superconductivity of the crystal in zero magnetic field calorimetrically as $T_{c0} \approx 16.8\, $K. The crystal undergoes a martensitic phase transition from cubic to tetragonal upon cooling through a temperature $T_M \approx 16.7\, $K (see Fig.~\ref{fig.physicaC1} and later Fig.~\ref{fig.physicaC7}) as soon as superconductivity is suppressed by a magnetic field. Fig.~\ref{fig.physicaC21} shows the phase diagram of the investigated sample, showing the upper critical field $H_{c2}(T)$ and the weakly field dependent martensitic transition at $H_M(T)$.\\
\indent The LuNi$_2$B$_2$C single crystal investigated here ($m \approx 13.2\, $mg) was irregularly shaped, with only a few smooth surfaces. No previous data of physical properties were available for this crystal. We determined the transition temperature as $T_{c0} \approx 15.5\, $K. In Fig.~\ref{fig.physicaC20} we show the phase diagram of this sample, with an unusual positive curvature of $H_{c2}(T)$ near $T_{c0}$. This is an indication for a high quality of the crystal in which the clean limit is achieved \cite{Drechsler1999}.\\
\indent The $2$H-NbSe$_2$ sample under study ($m \approx 6.4\, $mg) consisted of $6$ thin slices cut from a large single crystal, which we stuck together with Apiezon N grease. 
The sample appeared to be somewhat inhomogeneous since we observed two superconducting phase transitions in a narrow temperature interval around $T = 6.8\, $K, separated by $\approx 80\, $mK (see below). This inhomogeneity might be a property of the crystal or was later induced by the cutting process. We show results on this ensemble of crystals to demonstrate that high-resolution specific-heat data can be used to clearly identify different superconducting phases even if the respective critical temperatures are very close to each other.

\section{Results}

\subsection{1. Phase transition in V$_3$Si, LuNi$_2$B$_2$C, and $2$H-NbSe$_2$}

\subsubsection{a) V$_3$Si}

\begin{figure}
\includegraphics[width=75mm,totalheight=200mm,keepaspectratio]{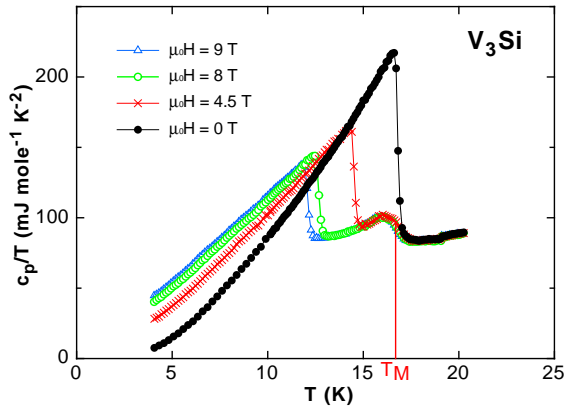}
\caption{Reduced specific heat $c_p/T$ of a V$_3$Si single crystal for four different magnetic fields, measured in a commercial PPMS (Quantum Design) by a conventional heat-relaxation technique.
\label{fig.physicaC1}}
\end{figure}

\begin{figure}
\includegraphics[width=75mm,totalheight=200mm,keepaspectratio]{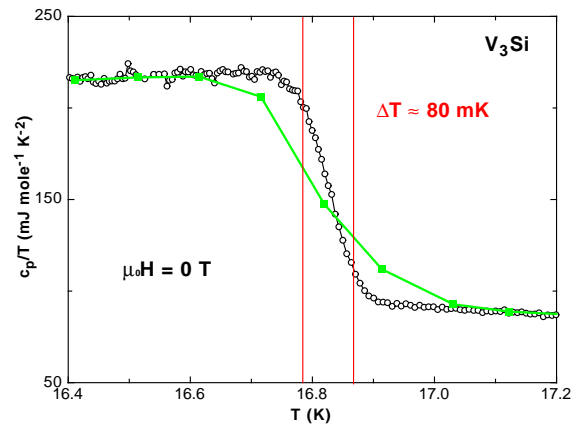}
\caption{Magnified view of $c_p/T$ of V$_3$Si in zero magnetic field in the transition region around $T_{c0}$. DTA data : open circles. PPMS data : squares. The approximate transition width ($10$-$90\%$ criterion) for the DTA measurement is marked with vertical lines.
\label{fig.physicaC2}}
\end{figure}

\begin{figure}
\includegraphics[width=75mm,totalheight=200mm,keepaspectratio]{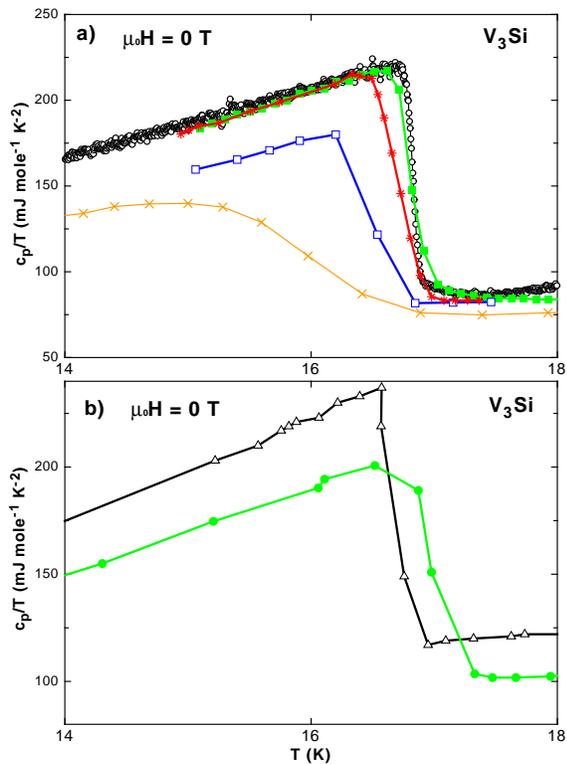}
\caption{Specific heat of V$_3$Si around $T_{c0}$ in zero magnetic field, measured on different samples with different techniques. \textbf{(a)} DTA method: single crystal, open circles (this work). Relaxation method: commercial PPMS, same crystal as DTA data, filled squares (this work); single crystal from \cite{Sebek2002February}, open squares; polycrystalline sample from \cite{Ramirez1996January}, crosses. Adiabatic method: single crystal from \cite{Khlopkin1999January}, stars. \textbf{(b)} Heat-pulse method: single crystal from \cite{Brock1969}, triangles; single crystal from \cite{Junod1980}, filled circles.
\label{fig.physicaC23}}
\end{figure}

\begin{figure}
\includegraphics[width=75mm,totalheight=200mm,keepaspectratio]{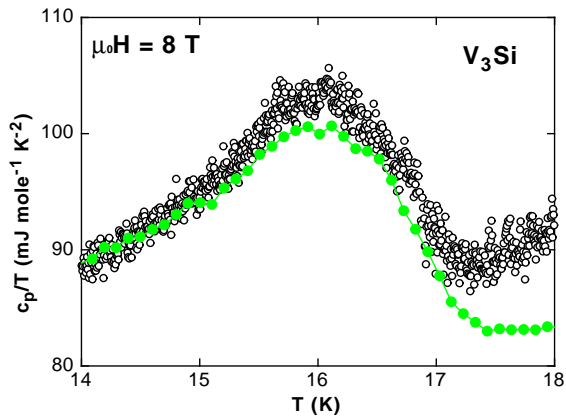}
\caption{Magnification of the martensitic transition region around $T_M$ of V$_3$Si. Filled circles: specific-heat data obtained with a commercial PPMS \textbf{relaxation method}. Open circles: DTA data on the same crystal measured in a magnetic field $\mu_0 H=8\, $T.
\label{fig.physicaC7}}
\end{figure}

\noindent In the first part of this section, we will compare our zero-field data acquired with the DTA method with data collected in a commercial PPMS (Quantum Design) by a \textbf{conventional heat-relaxation technique}. Fig.~\ref{fig.physicaC1} shows the temperature dependence of the specific heat of the V$_3$Si crystal for four different magnetic fields, measured with the DTA technique. While the transition to the superconducting state is strongly field dependent, the martensitic transition shows only a weak field dependence (see Fig.~\ref{fig.physicaC21}). The martensitic transition at $T_M$ takes place very close to the superconducting transition $T_{c0}$ in zero field and therefore causes a certain broadening of this transition.\\
\indent In Fig.~\ref{fig.physicaC2} we show both DTA and PPMS data in zero magnetic field at the transition region around $T_{c0}$, measured on the same V$_3$Si crystal. Applying a common ($10$-$90\%$)-criterion to the PPMS data, we find $\Delta T \approx 280\, $mK for the width $\Delta T$ of the phase transition. Near $T_{c0}$, the data-point density for this kind of measurement is $\approx 10$ data points per Kelvin, and the PPMS setup needed about $40\, $min to measure the displayed data. The reason for this relative low data-acquisition rate is that both the stabilization of the temperature and the subsequent relaxation for every single data point take considerable time.
In the same figure we show corresponding specific-heat data collected with the \textbf{DTA method} on the same sample. Near $T_{c0}$, the data-point density is $165$ data points per Kelvin, and the DTA experiment took about $8\, $min to collect these data. Using again the ($10$-$90\%$)-criterion we find $\Delta T \approx 80\, $mK for the width of the phase transition to the superconducting state. This transition is much narrower in the DTA data than in the relaxation method data from the PPMS (by a factor $\approx 3.4$). Since both measurements have been conducted on the same V$_3$Si crystal, this discrepancy in the transition width cannot be attributed to sample-property issues, but must be ascribed to the different measuring techniques. While every single temperature reading in a DTA experiment can represent, in principle, one $c_p$ data point \cite{Schilling1995October,Schilling2007March}, it is the temperature interval needed to measure the relaxation rate that determines the temperature resolution in a relaxation method experiment. This quantity is usually of the order of $100\, $mK or more, and can only be reduced at the cost of a larger uncertainty and scatter in the $c_p$ data.\\
\indent We next want to compare the DTA data to results of corresponding specific-heat measurements published by other groups for the same compound that have been obtained by different methods (see Fig.~\ref{fig.physicaC23}). It is obvious from Fig.~\ref{fig.physicaC23} that the difference in the respective transition widths is prominent. We have reproduced the data of Sebek \emph{et~al.}~\cite{Sebek2002February} in Fig.~\ref{fig.physicaC23}a (open squares) that have also been measured using a commercial PPMS system. Here, the data-point density is $\approx 4$ data points per Kelvin, roughly $40$ times less than in our DTA measurement. The transition width appears to be about $\Delta T \approx 520\, $mK, which is about $\approx 6$ times larger than the transition width measured with our DTA method. At the same time this width is also about twice as large as the transition width that we measured for our V$_3$Si crystal with the same relaxation method, which may indicate different sample qualities. One can safely state here that the relaxation methods achieve a considerable lower resolution on the temperature scale than what we routinely reach with a DTA method, while the data-acquisition rate is much higher with the latter method.\\
\indent Next, we compare our DTA data to corresponding data from a \textbf{standard semi-adiabatic heat-pulse technique} \cite{Ramirez1995February}. As an example we have reproduced the data of Ramirez \cite{Ramirez1996January}, taken on a polycrystalline V$_3$Si sample in zero magnetic field, in Fig.~\ref{fig.physicaC23}a (crosses). This sample was reported to have a $T_{c0}=16.5\, $K and a transition width of $\sim 8\%$ \cite{Ramirez1996January} (i.e., $\Delta T \approx 1220\, $mK), which is about $15$ times larger than the transition width that we measured with our DTA method on a different sample. Besides effects due to the different measuring techniques, this huge discrepancy can be attributed to the polycrystalline nature of the sample measured in Ref.~\cite{Ramirez1996January} that might lead to a certain inhomogeneity and a broadening of the transition to superconductivity. The data-point density of that experiment near $T_{c0}$ is $\approx 4$ data points per Kelvin.\\
\indent A similar \textbf{heat-pulse method} was also used by Junod \emph{et~al.}, who measured the specific heat with a classical heat-pulse calorimeter on a V$_3$Si single crystal ($m \approx 3.6\, $g) \cite{Junod1980}. The transition width can be estimated from the data reproduced in Fig.~\ref{fig.physicaC23}b to $\approx 430\, $mK, which is about five times larger than in our DTA data. The data-point density in this experiment is of the order of $5$ data points per Kelvin.\\
\indent As a last example for the \textbf{heat-pulse method}, we have plotted corresponding data obtained on a V$_3$Si single crystal by Brock \cite{Brock1969}. The transition width, taken from the data plotted in Fig.~\ref{fig.physicaC23}b, is $\approx 300\, $mK, which is about four times larger than in our experiment. The data-point density near $T_{c0}$ is of the order of $6$ data points per Kelvin.\\
\indent Finally, we compare our DTA data to an \textbf{adiabatic method} as used by Khlopkin \cite{Khlopkin1999January}. Details about the measuring technique are not given in Ref.~\cite{Khlopkin1999January} and are, to our knowledge, not published. We assume that the method is either a heat-pulse based technique or a variant of a continuous-heating technique \cite{Junod1979}. The corresponding data of a single crystal of V$_3$Si ($m=1.5\, $mg) in zero magnetic field are reproduced from Ref.~\cite{Khlopkin1999January} and are plotted in Fig.~\ref{fig.physicaC23}a (stars). The data-point density here is $\approx 14$ data points per Kelvin, with a transition width $\Delta T \approx 350\, $mK.\\
\indent The A$15$ superconductor V$_3$Si usually undergoes a martensitic phase transition. This crystallographic phase transition is known to be very sensitive to point-like or off-stoichiometric defects. Only high-quality samples with a resistivity ratio $\rho = R(300\, $K)/$R(18\, $K)$>25$ are expected to undergo such a transition that can be revealed by specific-heat measurements. A Peierls instability merely opens a gap on a portion of the Fermi surface, and the associated specific heat should behave qualitatively similar to that of a superconductor \cite{Maita1972}. The superconducting gap competes with the Peierls gap, however, and therefore superconductivity suppresses the martensitic transition as can be clearly seen from the absence of a feature coming from the martensitic transformation in our zero-field data shown in Fig.~\ref{fig.physicaC2}. While this transition usually takes place well above $T_{c0}$ in most V$_3$Si crystals, the martensitic transition in the here investigated crystal takes place slightly below $T_{c0}$.\\
\indent Fig.~\ref{fig.physicaC7} shows our specific-heat data of V$_3$Si for both the PPMS \textbf{relaxation method} and the corresponding DTA data in a magnetic field $\mu_0 H=8\, $T. The transition width of the structural phase transition is in both cases $\Delta T \approx 790\, $mK. This coincidence is expected in this case since instrumental broadening is most prominent for very sharp features, while the specific-heat discontinuity of the martensitic transition is too broad to be affected by such effects.

\subsubsection{b) LuNi$_2$B$_2$C}

\begin{figure}
\includegraphics[width=75mm,totalheight=200mm,keepaspectratio]{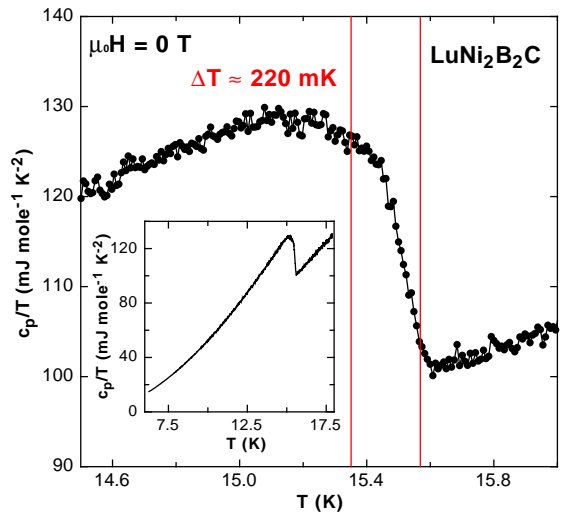}
\caption{Reduced specific heat $c_p/T$ in the transition region around $T_{c0}$ of a LuNi$_2$B$_2$C single crystal in zero magnetic field obtained with the DTA technique. The approximate transition width is marked with vertical lines. The inset shows the complete set of data.
\label{fig.physicaC8}}
\end{figure}

\begin{figure}
\includegraphics[width=75mm,totalheight=200mm,keepaspectratio]{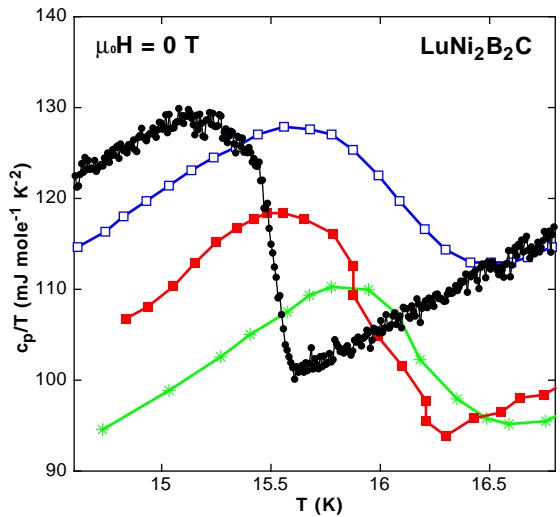}
\caption{Magnified view of the transition region around $T_{c0}$ for $c_p/T$ data obtained on different LuNi$_2$B$_2$C samples with different techniques in zero magnetic field. \textbf{DTA method}: single crystal, filled circles (this work). \textbf{Relaxation method}: polycrystalline sample \cite{Nohara1997July}, open squares. \textbf{Heat-pulse method}: polycrystalline sample \cite{Lipp2002May}, filled squares. The data of Kim \emph{et~al.}~\cite{Kim1994August} measured on a single crystal with an unknown technique is plotted with stars.
\label{fig.physicaC25}}
\end{figure}

Fig.~\ref{fig.physicaC8} shows the DTA data of a single crystal of LuNi$_2$B$_2$C ($m \approx 13.2\, $mg) in the transition region around $T_{c0}$ for zero magnetic field. The transition width is $\Delta T \approx 220\, $mK. The data-point density in that temperature region is $\approx 130$ data points per Kelvin. The inset shows the complete data set. Note the asymmetric rounding of the discontinuity at the transition to superconductivity at its low-temperature side, which somewhat complicates the determination of the transition width. At present we do not know whether this rounding is an intrinsic property of LuNi$_2$B$_2$C or if it is unique for our sample. Since this rounding is nevertheless a rather narrow feature ($\approx 200\, $mK wide), it might be smeared out in conventional measurements with lower resolution. However, if one does not attribute this feature to the transition to the superconducting state, the transition becomes much sharper, about $\approx 90\, $mK.\\
\indent In Fig.~\ref{fig.physicaC25} we have plotted the above discussed DTA data together with measurements on LuNi$_2$B$_2$C from the literature, which we will discuss in the following. As in the case of V$_3$Si, the difference in the respective transition widths is prominent.\\
\indent From the data of Nohara \emph{et~al.}~\cite{Nohara1997July} (open squares in Fig.~\ref{fig.physicaC25}), taken on a polycrystalline LuNi$_2$B$_2$C sample using a \textbf{relaxation method} \cite{Bachmann1972} we obtain, again using a ($10$-$90\%$)-criterion, a transition width $\Delta T \approx 840\, $mK, around nine times larger than we have measured on a single crystal using the DTA method. Again, this may be related to the polycrystalline nature of the sample investigated in Ref.~\cite{Nohara1997July}. Here, the data-point density is $\approx 10$ data points per Kelvin near $T_{c0}$.\\
\indent Corresponding data from another polycrystalline sample but measured using a \textbf{heat-pulse method} \cite{Lipp2002May} are also plotted in Fig.~\ref{fig.physicaC25} (filled squares). The transition width, $\Delta T \approx 420\, $mK, is approximately $4$-$5$ times larger than in our single-crystal DTA-data, which may again be related to sample-homogeneity issues. The data-point density in this work is $\approx 12$ data points per Kelvin.\\
\indent As a last example for LuNi$_2$B$_2$C, we refer to the data of Kim \emph{et~al.}~\cite{Kim1994August}, who investigated the specific heat of a single crystal of LuNi$_2$B$_2$C with an unknown method. From their data, reproduced in Fig.~\ref{fig.physicaC25} with stars, we deduce $\Delta T \approx 410\, $mK for the transition width, again $4$-$5$ times larger than the transition width measured with a DTA method. The data-point density in that work is $\approx 8$ data points per Kelvin.

\subsubsection{c) $2$H-NbSe$_2$}

\begin{figure}
\includegraphics[width=75mm,totalheight=200mm,keepaspectratio]{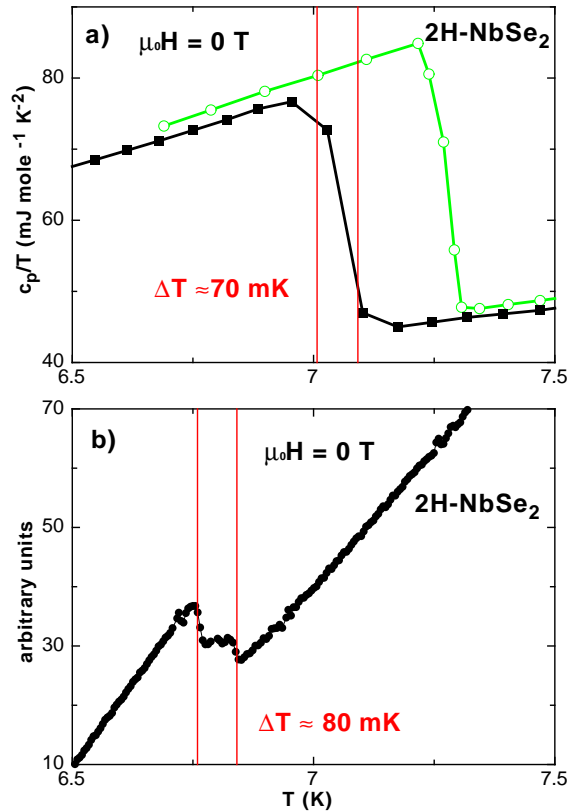}
\caption{Specific heat of $2$H-NbSe$_2$ in the transition region around $T_{c0}$ in zero magnetic field. We compare the data from Refs.~\cite{Sanchez1995,Nohara1999April} (a) with a corresponding DTA measurement on a stack of NbSe$_2$ crystals. (b) The two-step transition to superconductivity is clearly visible (see text). The respective transition widths are marked with vertical lines. \label{fig.physicaC26}}
\end{figure}

\begin{figure}
\includegraphics[width=75mm,totalheight=200mm,keepaspectratio]{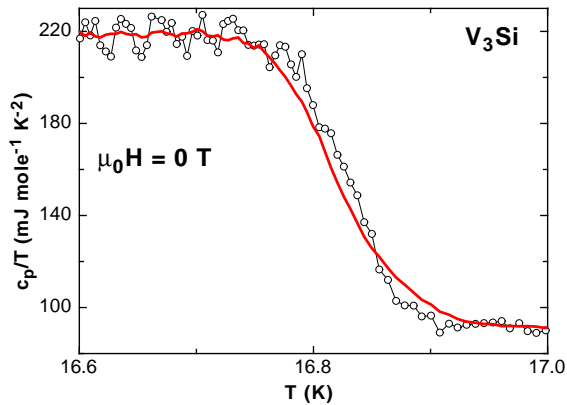}
\caption{Zero-field $c_p/T$ data on V$_3$Si obtained with the DTA method and evaluated with two different intervals $\delta T = 10\, $mK (open circles) and $\delta T \approx 110\, $mK (thick line).
\label{fig.physicaC28}}
\end{figure}

In this section, we will show that the high resolution achieved with the DTA method can be used to reveal possible sample inhomogeneities that would otherwise remain unnoticed. In Fig.~\ref{fig.physicaC26}a we reproduced the data of Sanchez \emph{et~al.}~\cite{Sanchez1995} (squares), obtained on a single crystal of $2$H-NbSe$_2$ with a \textbf{relaxation method} \cite{Sanchez1992}. The data-point density is about $15$ data points per Kelvin, and the transition width $\Delta T \approx 70\, $mK. In the same figure, we have plotted the data of Nohara \emph{et~al.}~\cite{Nohara1999April} obtained with a \textbf{relaxation calorimeter} \cite{Nohara1997July,Bachmann1972} for comparison (open circles). The data-point density here is again $\approx 15$ data points per Kelvin, and $\Delta T \approx 60\, $mK.\\
\indent In Fig.~\ref{fig.physicaC26}b we show corresponding DTA data that we have obtained on a NbSe$_2$ sample consisting of several single crystalline pieces stacked together with Apiezon N grease. The data-point density is about $190$ data points per Kelvin in the transition region. The DTA data clearly show two steps at the transition to superconductivity, which indicates the presence of areas with different transition temperatures. The transition width of the whole two-step transition is $\Delta T \approx 80\, $mK but we expect that the transition width of a homogeneous sample with only one $T_{c0}$ would be considerably smaller. This two-step transition is still rather sharp and can be clearly resolved with the DTA method, while it is questionable if such a feature could be resolved using a method with larger instrumental broadening and/or lower data-point density.\\
\indent To conclude this paragraph, we want to briefly comment on advantages and some peculiarities of the DTA technique as compared to the heat-pulse and relaxation techniques. Apart from advantages concerning speed and data-point density (see Table \ref{tab.summary}), the resolution of the DTA-$c_p$ data on the temperature scale can be adjusted \emph{after} a measurement has been done. The reason is that the DTA technique principally measures changes in entropy $S$ without any significant instrumental broadening effects \cite{Schilling2007March}. While in the other two techniques, the resolution in temperature is fixed by the finite temperature intervals used to obtain one single data point in $c_p$ (e.g., the pulse height or the temperature range in which a relaxation is measured), the resolution in temperature of $c_p(T)$ data obtained by the DTA technique is determined by numerical parameters that can be adjusted at will during data processing after the actual measurement. To be more precise, to obtain specific-heat $c_p/T$ data, the derivative of $S(T)$ has to be calculated numerically. This results in $c_p(T)$ data with an impressive resolution on the temperature scale. However, any attempt to increase the accuracy of $c_p$ on the temperature scale is at the expense of the accuracy in $c_p/T$ as it was discussed in detail in Refs.~\cite{Schilling1995October} and \cite{Schilling2007March}. The numerical calculation of $\delta S / \delta T$ from experimental data introduces a certain broadening $\delta T$ in $T$ given by the temperature interval used to calculate this derivative. Moreover, the scatter $\delta c_p$ in the specific-heat data as calculated from such a procedure \cite{Schilling2007March} is inevitably influenced by such a procedure. It has been shown that the product $\delta c_p \delta T$ is a constant that is essentially determined by the total heat capacity and by the limiting accuracy with which the temperatures can be measured \cite{Schilling1995October,Schilling2007March}. In other words, any attempt to increase the accuracy of $c_p$ in $T$, e.g., by choosing a narrower interval to calculate $\delta S / \delta T$, leads to an increased scatter $\delta c_p$. In order to illustrate this issue, we plotted in Fig.~\ref{fig.physicaC28} corresponding data evaluated with two different intervals $\delta T = 10\, $mK (open circles) and $\delta T \approx 110\, $mK (thick line) to numerically calculate $c_p/T$ from the same zero-field $S(T)$ data on V$_3$Si. It can clearly be seen that the increase in $\delta T$ (here by a factor $\approx 11$) leads to a reduction in scatter $\delta c_p$, but sharp features like the second order phase transition to superconductivity become more broad on the temperature scale. A detailed discussion of this issue can be found in Refs.~\cite{Schilling1995October,Schilling2007March}.\\
\indent While the present comparison has essentially been restricted to the very common heat-pulse and relaxation techniques to measure a specific heat, we want to mention that modern methods of advanced ac calorimetry and continuous-heating techniques can reach at least a comparable resolution on the temperature scale as DTA measurements. As soon as the temperature amplitude in an ac experiment is sufficiently small (as, e.g., in the work of Lortz \emph{et~al.}~\cite{Lortz2006September} using a thermopile-type arrangement), a temperature resolution of $1\, $mK is routinely achieved. As far as the data-acquisition time and the data-point density is concerned, continuous-heating methods \cite{Junod1979} share the same advantages with the DTA technique because both methods operate with a continuously varying temperature of the sample platform while monitoring the sample temperature as a function of time.

\begin{table}[h!]
\centering
\begin{tabular}{|l|c|c|c|c|c|c|}
\hline
\multicolumn{1}{|c|}{Method} & \multicolumn{1}{|c|}{Reference} & \multicolumn{1}{|c|}{Substance} & \multicolumn{1}{|c|}{Single crystal} & \multicolumn{1}{|c|}{$\Delta T$(mK)} & \multicolumn{1}{|c|}{$\rho_{dp}$(K$^{-1}$)} \\
\hline
DTA & This work & V$_3$Si & Yes & $80$ & $165$ \\
\hline
Relaxation & This work & V$_3$Si & Yes & $280$ & $10$ \\
\hline
Relaxation & \cite{Sebek2002February} & V$_3$Si & Yes & $520$ & $4$ \\
\hline
Heat-pulse & \cite{Ramirez1996January} & V$_3$Si & No & $1220$ & $4$ \\
\hline
Heat-pulse & \cite{Junod1980} & V$_3$Si & Yes & $430$ & $5$ \\
\hline
Heat-pulse & \cite{Brock1969} & V$_3$Si & Yes & $300$ & $6$ \\
\hline
Adiabatic & \cite{Khlopkin1999January} & V$_3$Si & Yes & $350$ & $14$ \\
\hline
DTA & This work & LuNi$_2$B$_2$C & Yes & $220$ & $130$ \\
\hline
Relaxation & \cite{Nohara1997July} & LuNi$_2$B$_2$C & No & $840$ & $10$ \\
\hline
Heat-pulse & \cite{Lipp2002May} & LuNi$_2$B$_2$C & No & $420$ & $12$ \\
\hline
Unknown & \cite{Kim1994August} & LuNi$_2$B$_2$C & Yes & $410$ & $8$ \\
\hline
DTA & This work & NbSe$_2$ & $6$ slices & $< 80$ & $190$ \\
\hline
Relaxation & \cite{Sanchez1995} & NbSe$_2$ & Yes & $70$ & $15$ \\
\hline
Relaxation & \cite{Nohara1999April} & NbSe$_2$ & Yes & $60$ & $15$ \\
\hline
\end{tabular}
\caption{Comparison of the ($10$-$90\%$)-transition widths $\Delta T$ of the transition to superconductivity measured with different techniques on various samples. $\rho_{dp}$ is the data-point density (in data points per Kelvin). The time needed to take one data point is $\approx 3\, $s with the DTA method, while we estimate it to be of the order of minutes for all the other techniques listed here. \label{tab.summary}}
\end{table}

\subsection{2. Low-temperature magnetic-field dependence of the specific heats of V$_3$Si and LuNi$_2$B$_2$C}

\begin{figure}
\includegraphics[width=75mm,totalheight=200mm,keepaspectratio]{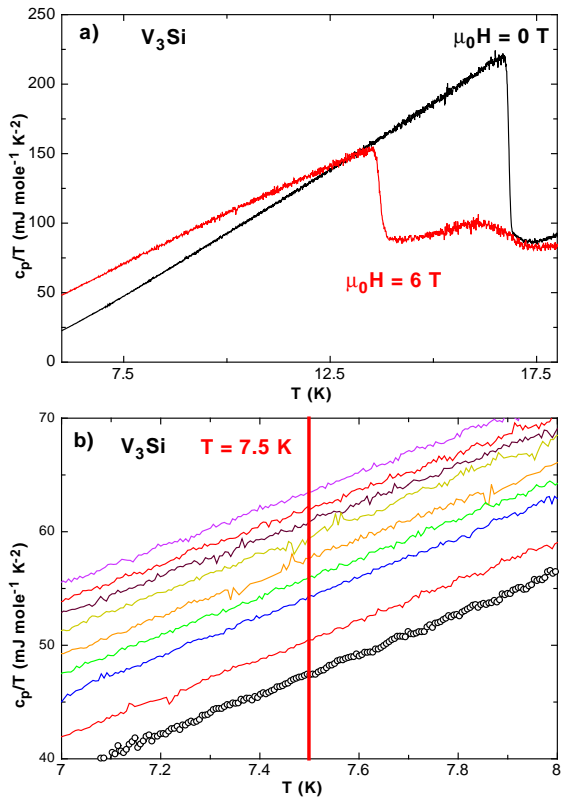}
\caption{$c_p/T$-DTA data of a V$_3$Si single crystal : \textbf{(a)} for $\mu_0 H=0\, $T and $6\, $T; \textbf{(b)} in a small temperature interval around $T=7.5\, $K, starting with $\mu_0 H=0\, $T (open circles) for the lowest curve and $\mu_0 H=4\, $T for the uppermost curve, $H$ increasing in intervals of $0.5\, $T.
\label{fig.physicaC40}}
\end{figure}

\begin{figure}
\includegraphics[width=75mm,totalheight=200mm,keepaspectratio]{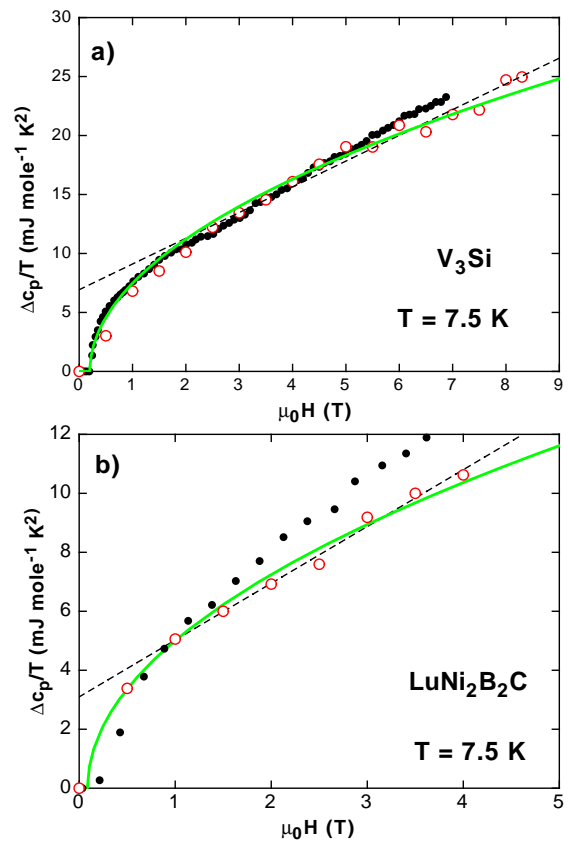}
\caption{Magnetic-field dependence of the specific-heat difference $\Delta c_p/T(H)$ at $T=7.5\, $K for different crystals. The thick lines represent fits according to $\Delta c_p/T(H)=A(H-H_{c1}^*)^{1/2}$ (see text). The dashed lines are guides to the eye to visualize the approximate linear behavior for the magnetic fields $\mu_0 H > 2\, $T in the case of V$_3$Si and $\mu_0 H > 1\, $T in the case of LuNi$_2$B$_2$C. \textbf{(a)} Data of a V$_3$Si single crystal obtained from DTA measurements (open circles), in comparison with the data of Ramirez \cite{Ramirez1996January} (filled circles). \textbf{(b)} Data of a LuNi$_2$B$_2$C single crystal obtained from DTA measurements (open circles), together with the data of Nohara \emph{et~al.}~\cite{Nohara1997July} (filled circles).
\label{fig.physicaC27}}
\end{figure}

\begin{figure}
\includegraphics[width=75mm,totalheight=200mm,keepaspectratio]{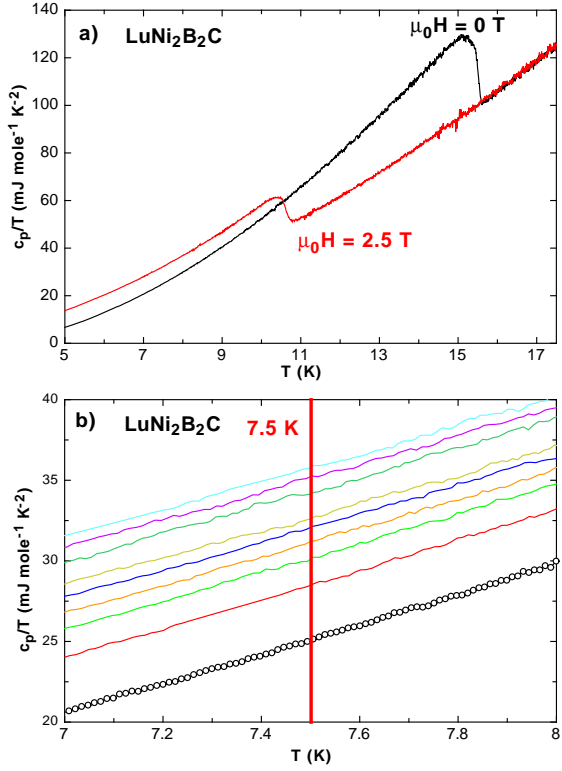}
\caption{$c_p/T$-DTA data of a LuNi$_2$B$_2$C single crystal: \textbf{(a)} for $\mu_0 H=0\, $T and $2.5\, $T; \textbf{(b)} in a small temperature interval around $T=7.5\, $K, starting with $\mu_0 H=0\, $T (open circles) for the lowest curve and $\mu_0 H=4\, $T for the upper most curve, $H$ increasing in intervals of $0.5\, $T.
\label{fig.physicaC41}}
\end{figure}

\begin{figure}
\centering
\includegraphics[width=75mm,totalheight=200mm,keepaspectratio]{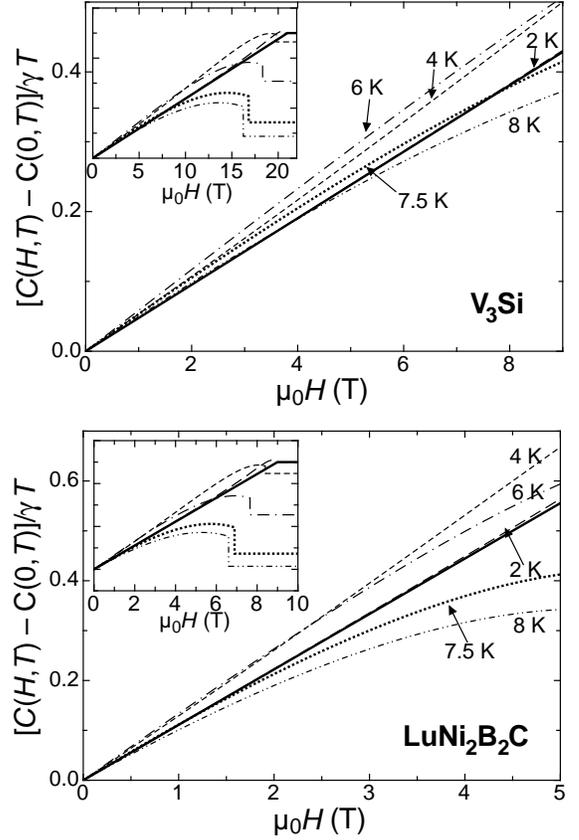}
\caption{Normalized differences $\Delta c_p/\gamma_NT(H)$ as modeled for V$_3$Si (upper panel) and  LuNi$_2$B$_2$C (lower panel) for different temperatures $T$, assuming a field dependent mixture of normal and superconducting components that vary proportional to $H/H_{c2}$ and $1-H/H_{c2}$, respectively (see text). Thick solid line: data for $T =0$.
\label{fig.14}}
\end{figure}

In the last section of this paper we show that specific-heat data from DTA measurements can reveal trends for the absolute value of $c_p$ even in the absence of a sharp phase transition. As an example, we focus on the low-temperature specific heats of V$_3$Si and LuNi$_2$B$_2$C that smoothly increase with the applied magnetic field $H$, but still may contain valuable information about the physics of the mixed state of these compounds.\\
\indent In the standard theory of superconductivity \cite{Gorter1964,Maki1965,Abrikosov1988} the magnetic-field dependence of the specific heat in the mixed state is nearly linear in $H$ for $T \rightarrow 0$ (i.e., proportional to the number of vortices), and can be written as

\begin{equation}
\Delta c_p/T(H)=(c_p(H)-c_p(H=0))/T=\gamma_N H/H_{c2}(T) \quad ,
\label{eq.1}
\end{equation}

\noindent where $\gamma_N$ is the Sommerfeld coefficient characterizing the specific heat of the electronic system in the normal state. This increase in $\Delta c_p/T(H)$ can, in the simplest picture, be ascribed to the contribution of the normal cores of the flux lines entering the sample, and a low-temperature linear term is expected to be proportional to $H$ up to $H_{c2}$ \cite{Caroli1965}. At temperatures $T > 0$, the field dependence of $\Delta c_p/T(H)$ is expected to be more complicated, and a non-linear behavior has been observed in a number of compounds: V$_3$Si \cite{Ramirez1996January}, LuNi$_2$B$_2$C \cite{Nohara1997July, Nohara1999April}, YNi$_2$B$_2$C \cite{Nohara1999April,Michor1995}, NbSe$_2$ \cite{Nohara1999April,Sanchez1995}, UPt$_3$ \cite{Ramirez1995February}, CeRu$_2$ \cite{Hedo1998January}, Lu$_2$Fe$_3$Si$_5$ \cite{Nakajima2008April}, and in YBa$_2$Cu$_3$O$_7$ \cite{Moler1994November,Wright1999February}.\\
\indent We have measured the specific heat of V$_3$Si in magnetic fields up to $\mu_0 H=8.3\, $T. Fig.~\ref{fig.physicaC40}a shows two DTA data sets for $\mu_0 H=0\, $T and $6\, $T. Below the transition to superconductivity and the martensitic transition the two data sets do not match, which is expected due to the field dependent $\Delta c_p/T(H)$. In Fig.~\ref{fig.physicaC40}b we show corresponding data in a small temperature interval around $T=7.5\, $K for selected magnetic fields. The non-linear field dependence at the lowest investigated magnetic fields is already visible in this figure.\\
\indent Extracting the $c_p/T(H)$ data for a fixed temperature $T = 7.5\, $K results in the data plotted in Fig.~\ref{fig.physicaC27}a, demonstrating the magnetic-field dependent specific-heat difference $\Delta c_p/T(H)$ of V$_3$Si (open circles). The dashed line is a guide to the eye and visualizes an approximately linear behavior above $\mu_0 H > 2\, $T. It can clearly be seen that the data deviate from linearity below $\mu_0 H < 2\, $T where they are better described by a $H^{1/2}$ dependence. For comparison, we have plotted in the same figure the data of Ramirez \cite{Ramirez1996January} from a non-transforming polycrystalline V$_3$Si sample, i.e., from a sample that does not exhibit martensitic transformation. The deviation from linearity becomes significant for low fields close to the lower critical field $H_{c1}$. In analogy to Ref.~\cite{Ramirez1996January} we can fit our data to $\Delta c_p/T(H) = A(H-H_{c1}^*)^{1/2}$ with $H_{c1}^*=0.2\, $T being the field where magnetic flux starts to enter the sample \cite{Swartz1962,Hauser1962}, and $A \approx 8.4\, $mJ mole$^{-1}$ K$^{-2}$T$^{-1/2}$, in good agreement with the result of Ref.~\cite{Ramirez1996January}.\\
\indent In Fig.~\ref{fig.physicaC41} we show the magnetic field dependent $c_p/T$ data for LuNi$_2$B$_2$C in a similar way as we did for V$_3$Si in Fig.~\ref{fig.physicaC40}. From these data we obtain again the specific-heat difference $\Delta c_p/T(H)$ at a fixed $T=7.5\, $K. It can clearly be seen in Fig.~\ref{fig.physicaC27}b that the data deviate from linearity below $\mu_0 H \approx 1\, $T where they again may be better described by a $H^{1/2}$ dependence. A tentative fit according to $\Delta c_p/T(H) = A(H-H_{c1})^{1/2}$ with $H_{c1} = 87\, $mT \cite{Schmiedeshoff2001March} gives $A \approx 5.2\, $mJ mole$^{-1}$ K$^{-2}$T$^{-1/2}$.\\
\indent For comparison we have reproduced the data of Nohara \emph{et~al.}~for LuNi$_2$B$_2$C in Fig.~\ref{fig.physicaC27}b. Our data clearly confirm the observation that also in LuNi$_2$B$_2$C $\Delta c_p/T(H)$ is not proportional to $H$.\\
\indent The prediction of linearity of $\Delta c_p/T(H)$ in $H$ strictly holds, within the BCS theory, only for $T \rightarrow 0$. To test whether or not the BCS prediction for the behavior of $\Delta c_p/T(H)$ data taken at fixed but \emph{elevated} temperatures might also account for the observed $\sqrt{H}$-like dependence of $\Delta c_p/T(H)$ we have simulated corresponding data according to a model that assumes a mixture of a normal-state electronic specific heat $c_{en}$ that is explicitly proportional to $H/H_{c2}$, and a superconducting component $c_{es}$ proportional to $(1-H/H_{c2})$. For the normal-state electronic specific heat we used $c_{en}=\gamma_NT$ and for the superconducting component a standard BCS $s$-wave expression $c_{es} = 9.17 \gamma_N T_c exp(-1.5 T_c/T)$, which is a fair approximation not too close to $T_c$ or $H_{c2}$, respectively. Taking our measured values for $T_{c0}$ and $T_c(H)=T_{c0}(1-H/H_{c2})^{1/2}$ with $\mu_0 H_{c2} = 21\, $T for V$_3$Si and $9\, $T for LuNi$_2$B$_2$C, we obtain the differences $\Delta c_p/T(H)$ as plotted in Fig.~\ref{fig.14}. It can clearly be seen that although the data for $T = 7.5\, $K are somewhat rounded as $H$ approaches the upper critical field $H_{c2}(T)$, the distinct peculiar curvature as observed in the corresponding experimental data for $H \rightarrow 0$ (see Fig.~\ref{fig.physicaC41}) can in no way be reproduced. A systematic study of Nohara \emph{et~al.}~\cite{Nohara1997July} on LuNi$_2$B$_2$C has indeed shown that the $\sqrt{H}$ dependence of $\Delta c_p/T(H)$ for $T \rightarrow 0$ is qualitatively preserved up to $T \approx 9\, $K.\\
\indent These findings are noteworthy since both V$_3$Si and LuNi$_2$B$_2$C are supposed to be \emph{s}-wave systems \cite{Weger1973,Orlando1981,Junod1983}. The deviation of $\Delta c_p/T(H)$ from linearity and especially its approximate $H^{1/2}$ dependence is very often interpreted as an indication for lines of nodes in the energy gap and taken as a hallmark for a \emph{d}-wave symmetry of the order parameter \cite{Wright1999February,Moler1994November,Volovik1993}. We can state here that a non-linear $\Delta c_p/T(H)$ as one observes in the \emph{s}-wave superconductors V$_3$Si and LuNi$_2$B$_2$C is likely a more general phenomenon of type II superconductors. Possible plausible explanations for this peculiar behavior may include vortex-vortex interactions \cite{Fetter1969,Maki1965,Ramirez1996January}, the field dependent shrinking of the vortex cores \cite{Nohara1999April,Kogan2005April}, and multigap scenarios \cite{Guritanu2004November,Shulga1998February}.\\
\indent In summary we have presented DTA-specific-heat measurements on various superconductors and compared the achieved temperature resolution to corresponding data obtained by other methods from the literature. The DTA method allows for an exceptionally high resolution on the temperature scale while providing a very high data-point density and a short measuring time. We have also investigated the magnetic-field dependence of the specific heats of V$_3$Si and LuNi$_2$B$_2$C at a fixed temperature $T=7.5\, $K, and found an excellent agreement with corresponding data from the literature. These measurements demonstrate that low-temperature DTA techniques are not only suitable to detect sharp phase transitions, but also allow for sufficiently precise measurements of absolute values of $c_p$.

\section{Acknowledgements}

We thank to G.~Krauss for technical assistance. This work was supported by the Schweizerische Nationalfonds zur F\"{o}rderung der Wissenschaftlichen Forschung, Grant.~No.~$20-111653$. Work at the Ames Laboratory was supported by the Department of Energy, Basic Energy Sciences, under Contract No.~DE-AC$02$-$07$CH$11358$.



\end{document}